\documentclass[runningheads]{llncs}
\usepackage[T1]{fontenc}
\usepackage{graphicx}
\usepackage{comment}
\usepackage{amsmath,amssymb} 
\usepackage{color}
\usepackage{url}
\usepackage[nocompress]{cite}

\newcommand{\bR}{\mathbb{R}}
\begin{document}
\title{Swin SMT: Global Sequential Modeling for Enhancing 3D Medical Image Segmentation}
\titlerunning{Swin SMT: Global Sequential Modeling in 3D Medical Image Segmentation}
%
\author{Szymon Płotka\inst{1,2} \and
Maciej Chrabaszcz\inst{2} \and
Przemyslaw Biecek\inst{2}}
%
\institute{
Informatics Institute, University of Amsterdam, Amsterdam, The Netherlands \\ 
\and
Warsaw University of Technology, Warsaw, Poland\\
\email{maciej.chrabaszcz.dokt@pw.edu.pl}
}
\authorrunning{S. Płotka et al.}

\newcommand\blfootnote[1]{%
  \begingroup
  \renewcommand\thefootnote{}\footnote{#1}%
  \addtocounter{footnote}{-1}%
  \endgroup
}

\maketitle              
\blfootnote{S. Płotka and M. Chrabaszcz -- Authors contributed equally.}
\begin{abstract}

Recent advances in Vision Transformers (ViTs) have significantly enhanced medical image segmentation by facilitating the learning of global relationships. However, these methods face a notable challenge in capturing diverse local and global long-range sequential feature representations, particularly evident in whole-body CT (WBCT) scans. To overcome this limitation, we introduce Swin Soft Mixture Transformer (Swin SMT), a novel architecture based on Swin UNETR. This model incorporates a Soft Mixture-of-Experts (Soft MoE) to effectively handle complex and diverse long-range dependencies. The use of Soft MoE allows for scaling up model parameters maintaining a balance between computational complexity and segmentation performance in both training and inference modes. We evaluate Swin SMT on the publicly available TotalSegmentator-V2 dataset, which includes 117 major anatomical structures in WBCT images. Comprehensive experimental results demonstrate that Swin SMT outperforms several state-of-the-art methods in 3D anatomical structure segmentation, achieving an average Dice Similarity Coefficient of 85.09\%. The code and pre-trained weights of Swin SMT are publicly available at \url{https://github.com/MI2DataLab/SwinSMT}.

\end{abstract}

\section{Introduction}

Segmentation of anatomical structures is crucial for various clinical applications such as diagnostics, treatment planning, quantitative analysis, image-guided interventions, and clinical trials~\cite{litjens2017survey}. In recent years, various segmentation methods of anatomical structures have been developed~\cite{he2023swinunetr,hatamizadeh2022unetr}, particularly for specific tasks such as organ segmentation and tumor delineation~\cite{heller2020state,bilic2023liver}, as well as for specific body parts like abdominal organs~\cite{ma2021abdomenct,luo2022word,ji2022amos}. 

\begin{figure}[t!]
    \centering
    \includegraphics[width=\textwidth]{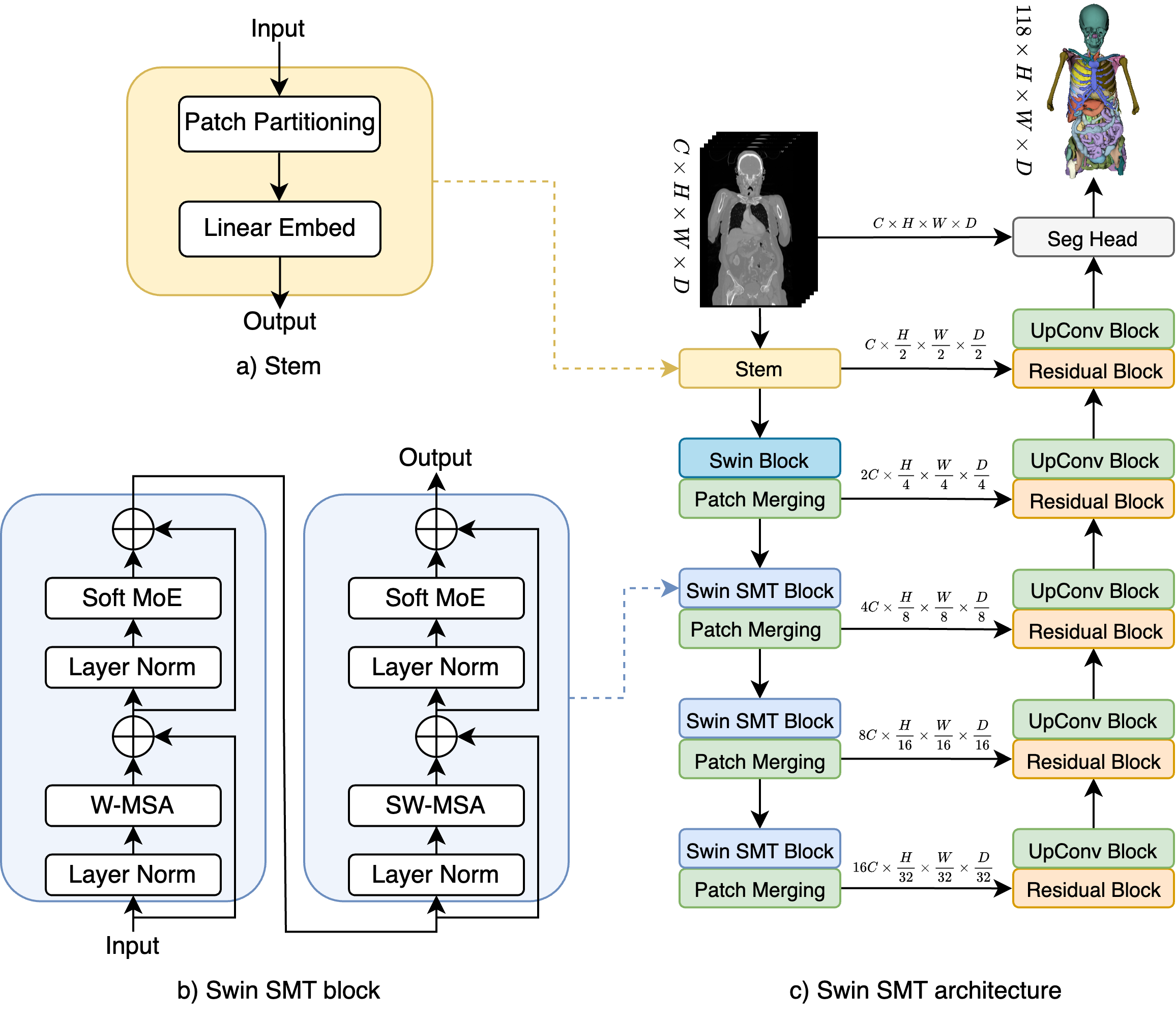}
    \caption{An overview of the Swin SMT architecture. The input to our model is a 3D CT scan. The Stem creates non-overlapping patches of the input data and utilizes a patch partition layer to generate windows with a desired size for computing self-attention. Encoded feature representations in each of the encoder blocks are then fed to a Convolutional Neural Network (CNN)-based decoder via skip connections at multiple resolutions. The segmentation output consists of 118 channels, corresponding to 117 classes and background, representing the major anatomical structures in WBCT images. H, W, D, and C refer to height, width, depth, and number of feature channels, respectively. W-MSA and SW-MSA refer to window-based multi-head self-attention with regular and shifted windows, respectively.}
    \label{fig:swinsmt_overview}
\end{figure}

While large-scale ViT-based methods have garnered significant attention for their impressive performance and capacity to model long-range sequences, they encounter notable challenges when employed in medical image analysis. These challenges include computational complexity, as seen in architectures such as ViT-based backbone (e.g. UNETR~\cite{hatamizadeh2022unetr}), RepUX-Net~\cite{lee2023scaling}, or nnFormer~\cite{zhou2023nnformer}, and a deficiency in modeling long-range dependencies in scale and intensity, as observed in models like Swin UNETR~\cite{tang2022self} or UNesT~\cite{yu2023unest}.

In recent years, Mixture-of-Experts (MoE)~\cite{Lepikhin2020GShardSG} has gained popularity in the Natural Language Processing domain and has subsequently been applied in the computer vision domain. MoE allows the model to dynamically allocate computational resources to the most relevant parts of the input data, enhancing efficiency and performance. By selectively activating different subset of the model's parameters, MoE enables the ViT to scale up its capacity without a corresponding increase in computational costs, leading to more accurate and efficient processing of complex visual information~\cite{Lepikhin2020GShardSG,Puigcerver2023FromST}. Unlike vanilla MoE, recently introduced Soft MoE~\cite{Puigcerver2023FromST} aggregates sequences into multiple global representations by using a weighted average over sequence elements, enabling the model to select the most important features. In the case of WBCT scans, global representations are crucial for the proper segmentation of certain classes with various scales and intensities, such as ribs and vertebrae.

Following these findings, we propose a novel \textbf{Swin} \textbf{S}oft \textbf{M}ixture \textbf{T}ransformer (Swin SMT), based on the Swin UNETR architecture, that incorporates Soft MoE to effectively handle complex and diverse long-range dependencies, as seen in WBCT images. Soft MoE allows for scaling up the model parameters maintaining a balance between computational complexity and segmentation performance in both training and inference modes. Swin SMT is designed for modeling the whole volume global features at various scales and intensities. To the best of our knowledge, this marks the first attempt to employ Soft MoE for the segmentation task. It is also the first benchmark of the publicly available TotalSegmentator-V2 dataset \cite{wasserthal2023totalsegmentator}, encompassing 117 major anatomical structures in WBCT scans.

\section{Method}
Our Swin SMT is built upon the Swin UNETR~\cite{tang2022self}, with a particular focus on enhancing the Transformer encoder. We incorporate the Soft MoE by replacing the feedforward network (FFN) within the Swin ViT block. The CNN-based decoder remains consistent with the original implementation~\cite{tang2022self}. An overview of the proposed method is illustrated in Fig.~\ref{fig:swinsmt_overview}.

\subsection{Swin Transformer Block}

Let the input $X\in \mathbb{R}^{C\times H \times W \times D}$ be a 3D CT scan. Then, Stem layer first partitions scan $X$ into a sequence of $2\times 2 \times 2$ scan patches returning $X'\in \mathbb{R}^{(HWD/8)\times 8C}$ and later applies a linear layer. With input $X$, the stem layer can be described as:
\begin{equation}\label{eq:stem}
    X' = \text{Partition}(X), \quad X_{\text{out}} = \text{Linear}(X'),
\end{equation}
where $X_{out}\in\mathbb{R}^{(HWD/8)\times d}$, and $d$ is the hidden size.

We use the Swin block at the first stage due to its reported computational and performance efficiency~\cite{tang2022self}. While Swin blocks excel at creating local representations, they encounter issues with long-range dependencies due to the windowed attention mechanism used in them, a challenge we aim to mitigate in our model with later stages.

\subsection{Swin Soft Mixture Transformer Block}

\begin{figure}[t!]
    \centering
    \includegraphics[width=\textwidth]{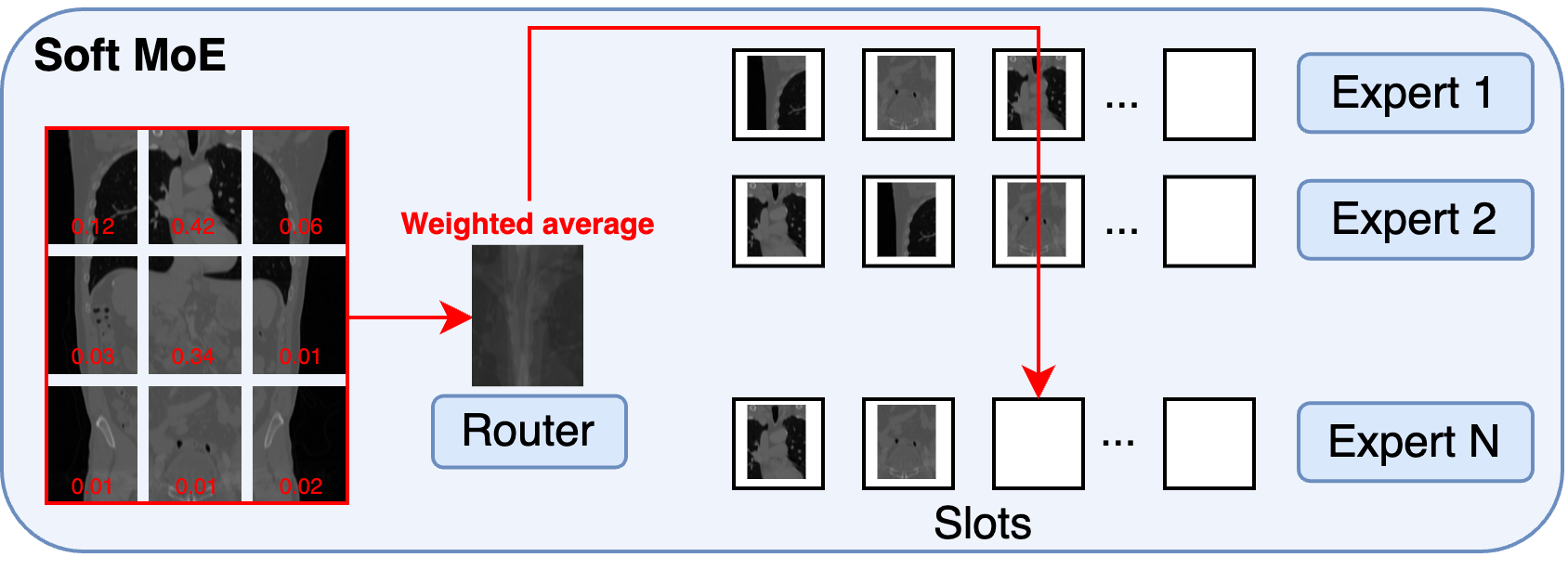}
    \caption{An overview of the Soft MoE. Here, the router assigns the weighted average of all the input tokens (patches) to each slot, which computes logits for each input pair of tokens and slots using dispatch weights. Then, each expert processes its slots. Finally, the original logits are normalized per token and used to combine all the slot outputs for every input token. Tokens in slots are shown in decreasing order of logits assigned to this token by dispatch weights.}
    \label{fig:softmore}
\end{figure}

Soft MoE calculates representations as convex combinations of tokens. This allows our model to better capture global information about WBCT scans by enabling each slot to create a comprehensive aggregation of input tokens. An overview of the Swin SMT block is presented in Fig.~\ref{fig:softmore}.

Motivated by~\cite{Puigcerver2023FromST}, we swap FFN with Soft MoE layers within the Swin Transformer block. Due to memory complexity, Soft MoE layers are added in each stage, except for the first stage. Let $X \in \bR^{m\times d}$ denote the input sequence, where $m$ is sequence length, and $d$ is hidden dimension size. With the model having $n$ experts and $s=m/n$ slots per expert, the memory complexity of Soft MoE is $\mathcal{O}(m^2 + md + nd^2)$. For the Swin block, we have $m=(p/2^i)^3$ at each stage, where $p$ is the patch size, and $i$ is the stage number ($i\in {1,2,3,4}$). Then, for $p=128$ at the first stage, we have $m=262,144$, leading to significant memory demand. Therefore, we choose to skip adding Soft MoE at the first stage.

The Soft MoE layer computes slot representations as convex combinations of input tokens and learnable parameters $\Phi$. The slot representations $\tilde{X}$ are computed according to the following equations:
\begin{align}
        D_{ij} &= \frac{\exp((X\Phi)_{ij})}{\sum_{i'=1}^m\exp((X\Phi)_{i'j})}, \quad \tilde{X} = D^TX.
\end{align}
After obtaining the slot representations $\tilde{X}$, $n$ expert functions are applied to each of the corresponding $s$ slots. The expert functions are modeled as FFNs that transform $\tilde{X}$. The transformation is expressed as:
\begin{equation}
    Y = concat(\{FFN_{\lfloor i/n \rfloor}(\tilde{X}_i); i\in\{1,\dots,ns\} \}),
\end{equation}
where $Y$ is the output matrix obtained by concatenating the results of the $n$ expert functions, $FFN_{\lfloor i/n \rfloor}(\tilde{X}_i)$ represents the output of the $\lfloor i/n \rfloor$-th expert function applied to the $i$-th slot representation $\tilde{X}_i$, $\text{concat}(\{\ldots\})$ signifies the concatenation of the outputs from all expert functions.

Having calculated $Y$, we now compute dispatch weights, creating a convex combination of expert representations $Y$ as follows:
\begin{align}
        S_{ij} &= \frac{\exp\left((X\Phi)_{ij}\right)}{\sum_{j'=1}^{n\cdot s}\exp((X\Phi)_{ij'})}, \quad X_{out} = SY.
\end{align}
\noindent
Here, $S_{ij}$ represents the dispatch weight for the 
$i$-th input sequence token and the $j$-th slot representation, and $X_{out}$ is the final output, obtained by multiplying the dispatch weights matrix $S$ with the expert representation matrix $Y$.

\subsection{Swin Soft Mixture Transformer}

\noindent
\textbf{Encoder.}
Our encoder module is based on Swin Transformer~\cite{liu2021swin} architecture with Soft MoE layers replacing FFN layers in Swin blocks at stages 2, 3, and 4. Having the input of previous layer $z_{l-1} \in \bR^{m\times d}$, the outputs for those stages are calculated as follows
\begin{align}
    \begin{split}
        \hat{z}_l &= \text{W-MSA}(\text{LN}(z_{l-1})) + z_{l-1}, \quad z_l = \text{Soft-MoE}(\text{LN}(\hat{z}_l)) + \hat{z}_l \\
        \hat{z}_{l+1} &= \text{SW-MSA}(\text{LN}(z_l)) + z_l, \quad z_{l+1} = \text{Soft-MoE}(\text{LN}(\hat{z}_{l+1})) + \hat{z}_{l+1},
    \end{split}
\end{align}
where LN refers to the layer-norm layer, W-MSA, and SW-MSA represent window-based multi-head self-attention with regular and shifted windows, respectively, following the original Swin ViT implementation.

\noindent
\textbf{Decoder.} Our decoder module is based on the same architecture as the decoder in the Swin UNETR model. Convolutional blocks are employed to extract outputs from each of the four blocks and the bottleneck. These extracted features are then upsampled using deconvolutional layers and concatenated with features from a higher resolution level. Finally, convolution with a 1 $\times$ 1 $\times$ 1 kernel is applied to map features to segmentation maps.

\section{Experiments and Results}

In this section, we describe the dataset used for training and evaluation of the Swin SMT. Additionally, we provide implementation details along with qualitative and quantitative results. Through an ablation study, we demonstrate the importance of the Swin SMT component that has been integrated into the Swin Transformer block.

\noindent
\textbf{Dataset.} To train and evaluate our method, we use the publicly available TotalSegmentator-V2 dataset~\cite{wasserthal2023totalsegmentator}. The dataset contains 1,228 CT scans with annotations for 117 major anatomical structures in WBCT images. It includes scans of the WBCT, as well as thoracic, abdominal, neck, and pelvic scans. Additionally, the dataset contains clipped sub-volumes corresponding to these body parts. We resample all scans to 1.5 $\times$ 1.5 $\times$ 1.5 mm$^{3}$ isotropic resolution. We follow the original data split~\cite{wasserthal2023totalsegmentator} that uses 1082, 57, and 89 cases for training, validation, and testing, respectively.

\noindent
\textbf{Implementation details.} We randomly crop a patch size of 128 $\times$ 128 $\times$ 128 from around the input CT scans. Training is conducted using the batch size of 1 per GPU and the AdamW optimizer with a warm-up cosine scheduler for 1000 epochs. As a loss function $L$, we use a sum of Dice and Cross-entropy loss defined as $L = L_{D} + \Lambda L_{CE},$ where $L_{D}$, $L_{CE}$ are Dice and Cross-entropy loss, respectively. A grid search optimization in the range [0.5, 1.0] was performed, which estimated optimal value $\Lambda$ = 1. An initial learning rate is set to 1 $\times$ 10$^{-4}$, and the weight decay is set to 1 $\times$ 10$^{-5}$. We adopt data augmentations of intensity, rotation, and scaling on the fly during training. For training, we use a DGX workstation equipped with 8 $\times$ NVIDIA A100 40GB GPUs, utilizing the Distribution Data Parallel methodology. The intensity of the Hounsfield Unit was clipped to the range [-1024, 1024] and linearly scaled to [0,1]. During the inference, we use a sliding window with an overlap of 0.5 and a Gaussian kernel. To compare the inference times, we use a scan with the dimensions of [300, 300, 422]. We implement Swin SMT using Python 3.9, PyTorch 2.1~\cite{paszke2019pytorch}, and MONAI 1.2.0~\cite{cardoso2022monai}. We evaluate segmentation performance using the Dice Similarity Coefficient (DSC). We perform a one-way analysis of variance (ANOVA) to assess the significance of differences among the performance metrics of Swin SMT segmentation performance. We use p $<$ 0.05 as a threshold for statistically significant differences.

\begin{table}[t!]
    \caption{Quantitative results comparing Swin SMT with several state-of-the-art methods on the test set of TotalSegmentator-V2 are presented. We compare model parameters (in millions, M), inference time (in seconds, s), and segmentation performance results using DSC (\%). Due to limited space, we divided the 117 classes into subgroups: organs, vertebrae, muscles, ribs, vessels, and the average of all these classes. The best-performing results are highlighted in bold. (*) denotes statistically significant differences between Swin SMT and compared state-of-the-art methods.}
    \resizebox{\textwidth}{!}{%
    \centering
    \begin{tabular}{@{}l|c|c|ccccc|lccccc@{}}
    \hline
    \textbf{Method} & \textbf{Params (M)} & \textbf{Time (s)} & \textbf{Organs $\uparrow$} & \textbf{Vertebrae $\uparrow$} & \textbf{Muscles $\uparrow$} & \textbf{Ribs $\uparrow$} & \textbf{Vessels $\uparrow$} & \textbf{Overall $\uparrow$}\\
    \hline
    UNETR \cite{hatamizadeh2022unetr} & 102.02 & 35 & 73.84 & 60.70 & 82.35 & 69.27 & 61.49 & 70.88 (*)\\
    SwinUNETR-S \cite{tang2022self} & 18.34 & 15 & 78.21 & 63.43 & 85.02 & 69.98 & 62.23 & 73.90 (*)\\
    nnFormer \cite{zhou2023nnformer} & 149.30 & 99 & 79.26 & 73.87  & 74.97 & 74.03 & 74.97 & 75.48 (*)\\
    DiNTS \cite{he2021dints} & 147.00 & 150 & 80.05 & 71.42 & 85.32 & 73.71 & 70.13 & 77.64 (*)\\
    UNesT \cite{yu2023unest} & 87.30 & 45 & 80.75 & 71.93 & 86.43 & 72.79 & 69.61 & 77.70 (*)\\
    3D UX-Net \cite{lee2023d} & 53.00 & 74 & 83.03 & 79.54 & 86.99 & 82.54 & 75.01 & 82.53 (*)\\
    SwinUNETR-B \cite{tang2022self} & 72.76 & 37 & 83.46 & 79.76 & 87.57 & 82.61 & 75.23 & 82.81 (*)\\
    nnU-Net \cite{isensee2021nnu} & 370.74 & 300 & 82.02 & 82.89 & 86.98 & 85.27 & 75.51 & 83.44 (*)\\
    SwinUNETR-L \cite{tang2022self} & 290.40 & 145 & 83.26 & 82.02 & 87.99 & 83.82 & 75.60 & 83.59 (*)\\
    3D RepUX-Net \cite{lee2023scaling} & 65.80 & 80 & 80.85 & 84.00 & 87.63 & 84.22 & 75.91 & 83.81 (*)\\
    Universal Model \cite{liu2023clip} & 62.25 & 39 & 82.25 & \textbf{84.46} & 87.58 & 86.49 & 76.11 & 84.02 (*)\\
    \hline
    \hline
    \textbf{Swin SMT (ours)} & 170.78 & 60 & \textbf{83.70} & 83.03 & \textbf{88.70} & \textbf{86.60} & \textbf{77.54} & \textbf{85.09}\\
    \end{tabular}
    }
    \label{tab:tsresults}
\end{table}

\noindent
\textbf{Quantitative results.} To assess the effectiveness and performance of Swin SMT, we conduct a comprehensive analysis with several state-of-the-art methods, i.e., nnU-Net~\cite{isensee2021nnu}, UNETR~\cite{hatamizadeh2022unetr}, 3D UX-Net \cite{lee2023d}, three configurations of Swin UNETR-based (S, B, and L) \cite{tang2022self}, 3D RepUX-Net~\cite{lee2023scaling}, nnFormer~\cite{zhou2023nnformer}, DiNTS~\cite{he2021dints}, UNesT~\cite{yu2023unest}, and Universal Model~\cite{liu2023clip} using the test set of TotalSegmentator-V2. To ensure a fair comparison, all models are trained under identical hardware settings. To enhance the performance parity among the compared methods, we employ the same hyperparameters, including optimizers, learning rates, and learning rate schedulers, following the configurations presented in their original works. Quantitative results for the TotalSegmentator's-V2 test set are presented in Table~\ref{tab:tsresults}. Our proposed Swin SMT demonstrated superior segmentation performance across all subgroups except for vertebrae. Nevertheless, the Swin SMT achieves a higher average DSC of 85.09\% across all 117 classes. During inference, our Swin SMT, with 170.78 million parameters, achieves a 60-second processing time. The one-way ANOVA revealed statistically significant differences between other state-of-the-art methods and Swin SMT (p $<$ 0.05).


\begin{figure}[t!]
    \centering
    \includegraphics[width=\textwidth]{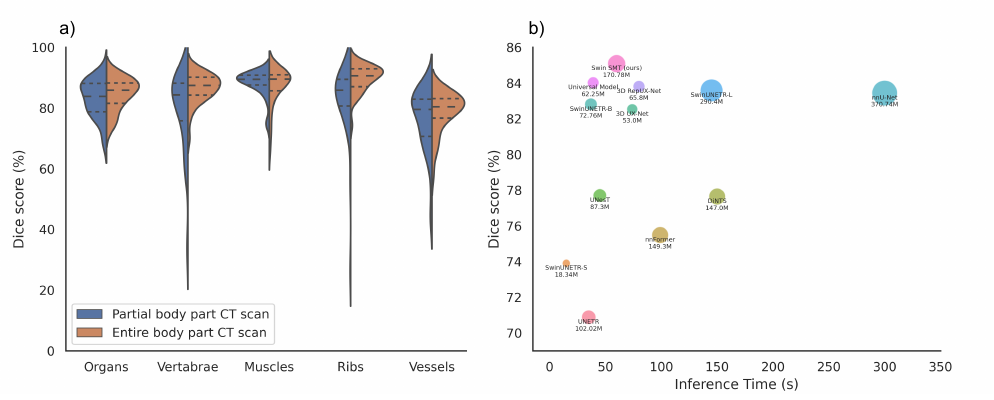}
    \caption{a) Distribution of the quantitative results of Swin SMT for each subgroup within WBCT images. The partial and full body part CT scans denote cropped (i.e., a sub-volume of a thoracic or abdominal) and entire body part (i.e., thoracic, abdominal, or whole-body) CT scans, b) Distribution of the average DSC against inference time (in s). The inference time calculations are based on an input patch of 128 $\times$ 128 $\times$ 128 with a sliding window algorithm and overlay of 0.5. The size of each circle indicates the number of parameters (in M).}
    \label{fig:violinplot}
\end{figure}

\noindent
\textbf{Ablation study.} We conduct an ablation study to investigate the impact of varying the number of experts on the segmentation performance of the Swin SMT. We train five configurations as follows. First, we train the baseline method -- Swin UNETR-B. Then, we swap FFN with Soft MoE in the Swin ViT block with 4, 8, 16, and 32 experts ($n$), respectively. Table~\ref{tab:ablationstudy} shows the results of an ablation study on the TotalSegmentator-V2 test set. The baseline method achieves a DSC of 82.81\%. The incorporation of the Soft MoE ($n=4$) into the baseline improved segmentation performance by 0.92\%. Further increasing the number of experts to 2$^{n}$, where $n\in {3, 4, 5}$ improved the DSC by 0.25, 0.56, and 1.36, respectively. The one-way ANOVA revealed statistically significant differences between the baseline and configurations with $n$ experts, where $n$ ranges from 4 to 32, including the best-performing configuration with $n=32$ experts (p $<$ 0.05). Due to limited computational resources, we did not use $>$ $n=32$ experts. 

\noindent
\textbf{Performance analysis.} Fig.~\ref{fig:violinplot}(a) shows the distribution of the quantitative results of Swin SMT for each subgroup of the WBCT images. We compare the evaluation on the test set of TotalSegmentator-V2, where \textit{Partial body part CT scan} and \textit{Entire body part CT scan} denote cropped (i.e., a sub-volume of a thoracic or abdominal) and entire body part (i.e., thoracic, abdominal, whole-body) CT scans, respectively. We can observe that the Swin SMT exhibits robust segmentation performance on entire-body scans. However, when it comes to sub-volumes, the model lacks contextual information and shows low performance on context-related classes such as vertebrae, ribs, and vessels. In the supplementary material, we provide qualitative results for both partial and entire-body scans. We present both top-performing cases for each subgroup and qualitative errors observed in WBCT scans, even when the quantitative results per scan (DSC above 90\%) are high.

\noindent
Fig.~\ref{fig:violinplot}(b) shows a speed-performance plot, indicating that Swin SMT presents a trade-off between computational efficiency and segmentation performance compared to several state-of-the-art methods while maintaining a high mean DSC score of 85.09\%. Unlike Swin SMT, which is 1.54 times slower than the Universal Model (2nd best), it is 1.33, 2.42, and 5.0 times faster than the next three best-performing methods, including 3D RepUX-Net, Swin UNETR-L, and nnU-Net, respectively.

\begin{table}[t!]
    \caption{The ablation study investigates the impact of varying the number of experts used in each Swin MoE block. We compare parameters (in millions, M), the number of experts (Experts), and DSC (\%). (*) denotes statistically significant differences between the top-performing configuration with n = 32 experts and all other configurations.}
    \begin{center}
        \begin{tabular}{@{}c|cclccccccccc@{}}
        \hline
        \textbf{Method} & \textbf{Params (M)} & \textbf{Experts (n)} & \textbf{DSC (\%)} $\uparrow$\\
        \hline
        Baseline & 62.19 & 0 & 82.81 (*)\\
        + Soft MoE ($n=4$) & 83.62 & 4 & 83.73 (*)\\
        + Soft MoE ($n=8$) & 87.01 & 8 & 83.98 (*)\\
        + Soft MoE ($n=16$) & 114.93 & 16 & 84.29 (*)\\
        + Soft MoE ($n=32$) & 170.78 & 32 & 85.09\\
        
        \end{tabular}
    \end{center}
    \label{tab:ablationstudy}
\end{table}

\section{Conclusions}

We proposed Swin SMT, a 3D Transformer-based method for segmenting major anatomical structures from WBCT images. By incorporating Soft MoE into the Swin Transformer block, Swin SMT effectively handles the complex and diverse long-range dependencies inherent in WBCT images. It is designed to model global features of the entire volume at various scales and intensities. A comprehensive benchmark, to the best of our knowledge, for the first time, on the publicly available TotalSegmentator-V2 dataset demonstrates superior performance compared to several state-of-the-art methods. Our method has the potential to serve as a support tool in clinical environments for the fast and accurate segmentation of major anatomical structures in WBCT images. In our future direction, we will incorporate large-scale self-supervised pretraining to enhance the performance of the segmentation model and evaluate its generalizability and transferability to external datasets.

\begin{credits}
\subsubsection{\ackname} This work was financially supported by the Polish National Center for Research and Development (grant no. INFOSTRATEG-I/0022/2021-00).

\subsubsection{\discintname} The authors have no competing interests to declare that are relevant to the content of this article. 
\end{credits}
%
%
%
%
\bibliographystyle{splncs04}
\bibliography{refs}

\newpage
\title{Supplementary material \\ Swin SMT: Global Sequential Modeling for Enhancing 3D Medical Image Segmentation}

\author{Szymon Płotka\inst{1,2} \and
Maciej Chrabaszcz\inst{2} \and
Przemyslaw Biecek\inst{2}}
%
\institute{
Informatics Institute, University of Amsterdam, Amsterdam, The Netherlands \\ 
\and
Warsaw University of Technology, Warsaw, Poland\\
\email{maciej.chrabaszcz.dokt@pw.edu.pl}
}
\authorrunning{S. Płotka et al.}

\maketitle

\begin{figure}[ht!]
    \centering
    \includegraphics[width=\textwidth]{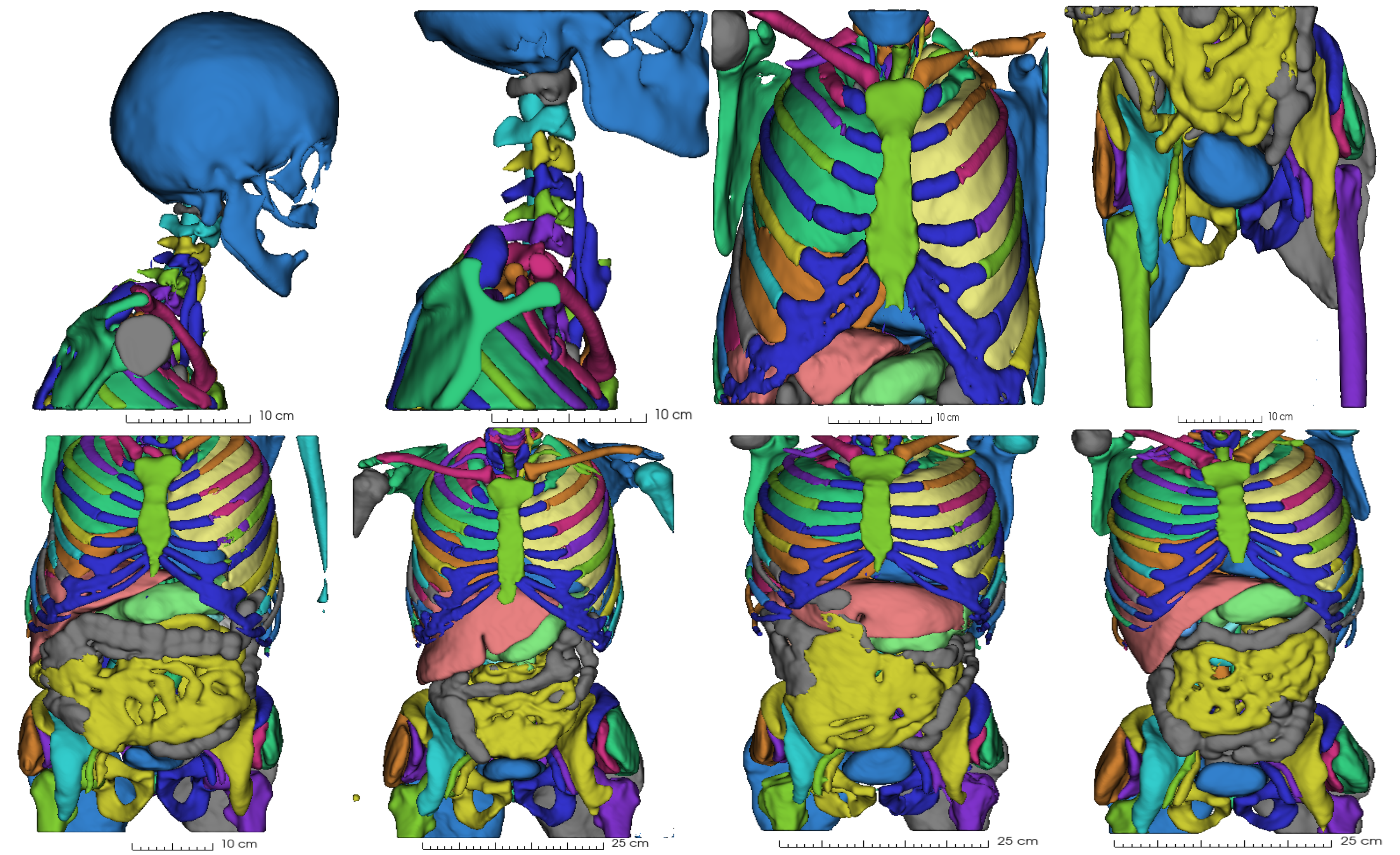}
    \caption{We show partial and entire-body part CT predictions of Swin SMT. On the top, we show partial predictions of various body parts, including head-neck, chest, and pelvis parts, which can provide less robust predictions due to less contextual information. On the bottom, we show entire-body predictions, for which model is more robust due to the global and contextual sequential information provided by Swin SMT. We achieved significantly higher segmentation performance on the entire-body scans rather than partial. We demonstrate our predictions with real word scale in centimeters [cm].}
    \label{fig:enter-label}
\end{figure}

\begin{figure}
    \centering
    \includegraphics[width=\textwidth]{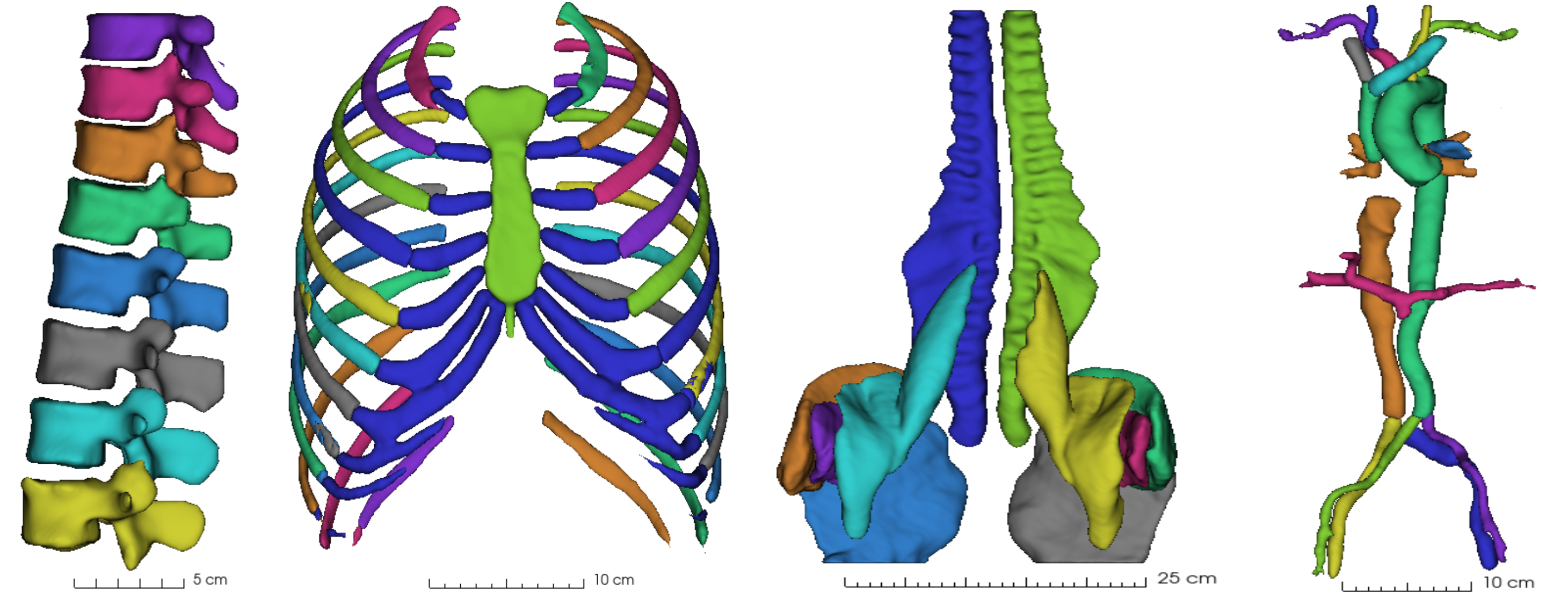}
    \caption{We show top-performing predictions of Swin SMT for various subparts of the entire-body predictions. We provide the Dice Score Coefficient (in \%) to show the segmentation performance. From the left: vertebrae (93.02\%), ribs with sternum and costal cartilages (95.07\%), muscles (92.34\%), and vessels (86.00\%). We demonstrate our predictions with real word scale in centimeters [cm].}
    \label{fig:enter-label}
\end{figure}

\begin{figure}
    \centering
    \includegraphics[width=\textwidth]{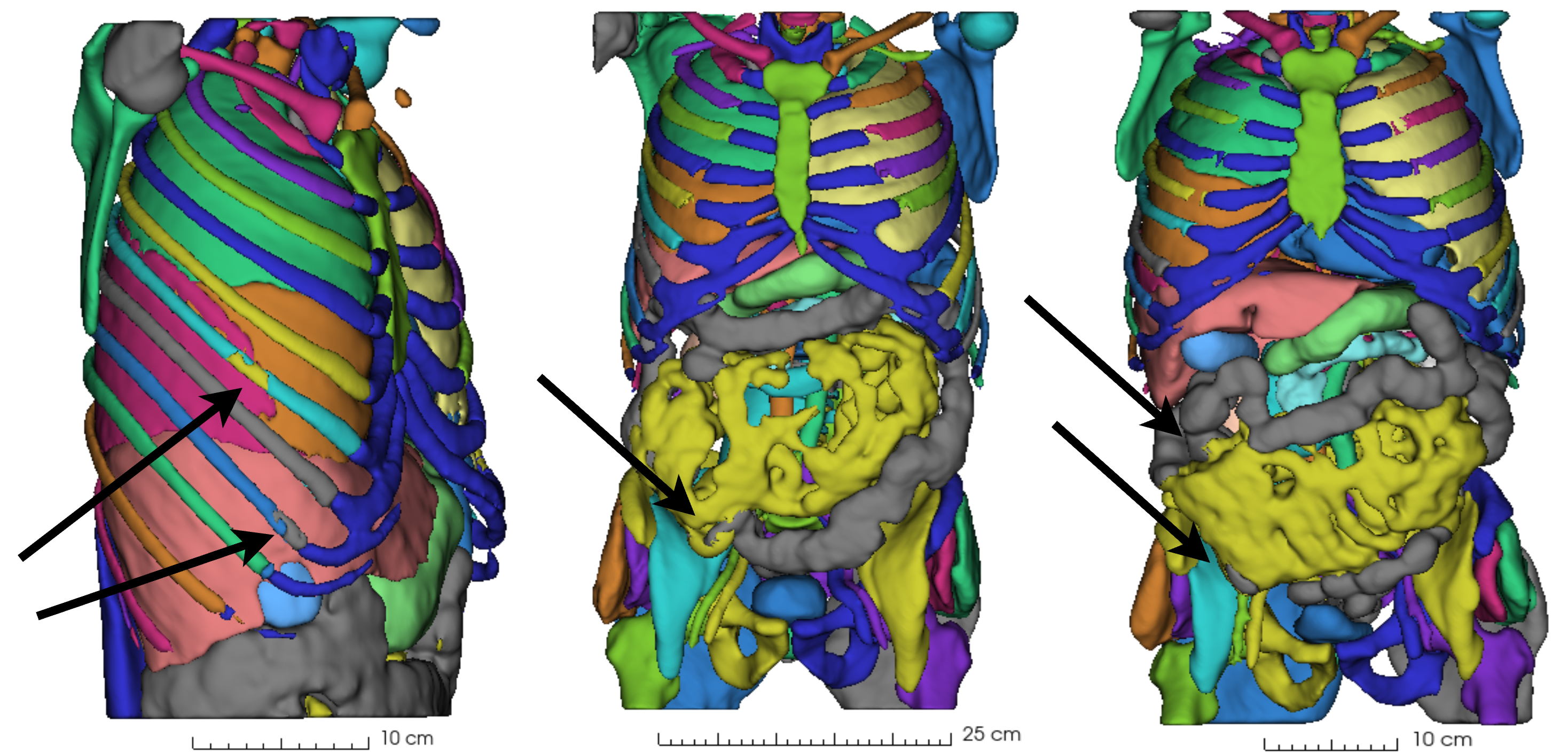}
    \caption{We show the errors of the predictions of Swin SMT. While the quantitative segmentation performance is high (above 90\% for each scan), we deal with some qualitative errors, as shown with black arrows. We demonstrate our predictions with real word scale in centimeters [cm].}
    \label{fig:enter-label}
\end{figure}

\end{document}